\input epsf
\documentstyle[12pt]{article}

        {\newpage
        \setcounter{page}{1}}

\renewcommand{\title}[1]{%
        {\begin{center}
        \Large\bf #1
        \end{center}}
        \vskip .3in}

\renewcommand{\author}[1]{%
        {\begin{center}
        #1
        \end{center}}}

\renewcommand{\abstract}[1]{%
        \begin{center}%
        {\vspace{1em}\vspace{0pt}\bf Abstract}%
        \end{center}%
        \noindent #1}

\renewcommand{\date}[1]{%
        \begin{center}%
        #1%
        \end{center}}

        {\end{thebibliography}}

\makeatletter
        \@addtoreset{equation}{section}%
\makeatother

\newcommand{\eqn}[1]{\label{eq:#1}}
\newcommand{\refeq}[1]{(\ref{eq:#1})}
\newcommand{\eq}{eq.~\refeq}
\newcommand{\Eq}{Eq.~\refeq}

\newcommand{\beq}{\begin{eqnarray}}
\newcommand{\eeq}{\end{eqnarray}}

\newcommand{\mybar}[1]%
        {\kern 0.8pt\overline{\kern -0.8pt#1\kern -0.8pt}\kern 0.8pt}
\newcommand{\tr}{{\rm tr}}
\def\twi{\widetilde}

\newcommand{\La}{\Lambda}
\newcommand{\LaN}{\Lambda_N}
\newcommand{\LaF}{\Lambda_F}
\newcommand{\LaFN}{\twi \Lambda_{\twi N}}

\newcommand{\drawsquare}[2]{\hbox{%
\rule{#2pt}{#1pt}\hskip-#2pt
\rule{#1pt}{#2pt}\hskip-#1pt
\rule[#1pt]{#1pt}{#2pt}}\rule[#1pt]{#2pt}{#2pt}\hskip-#2pt
\rule{#2pt}{#1pt}}

\newcommand{\Yfund}{\raisebox{-.5pt}{\drawsquare{6.5}{0.4}}}
\newcommand{\Ybarfund}{\mybar{\raisebox{-.5pt}{\drawsquare{6.5}{0.4}}}}%

\newcommand{\jref}[4]{{\it #1} {\bf #2}, #3 (#4)}
\newcommand{\NPB}[3]{\jref{Nucl.\ Phys.}{B#1}{#2}{#3}}
\newcommand{\PLB}[3]{\jref{Phys.\ Lett.}{#1B}{#2}{#3}}

\begin{document}

\begin{titlepage}
\begin{center}
{\hbox to\hsize{hep-th/9708015 \hfill  BUHEP-97-23}}

\bigskip
\bigskip
\bigskip
\vskip1in

{\Large \bf N = 1 Field Theory Duality from M-theory} \\

\bigskip
\bigskip
\bigskip
\vskip.4in

{\bf Martin Schmaltz and Raman Sundrum}\\

\bigskip
\bigskip

{\small \sl Department of Physics

Boston University

Boston, MA 02215, USA }

\smallskip

{\tt schmaltz@abel.bu.edu, sundrum@budoe.bu.edu}

\vspace{1.5cm}
{\bf Abstract}\\
\end{center}

\bigskip

We investigate Seiberg's $N=1$ field theory duality for
four-dimensional supersymmetric QCD with the M-theory 5-brane. We find
that the M-theory configuration for the magnetic dual theory arises via
a smooth deformation of the M-theory configuration for the electric
theory. The creation of Dirichlet 4-branes as Neveu-Schwarz 5-branes are
passed through each other in Type IIA string theory is given  a nice
derivation from M-theory.

\bigskip

\end{titlepage}

\section{Introduction}

In the last few years there has been tremendous progress in
understanding supersymmetric gauge dynamics and the remarkable
phenomenon of electric-magnetic duality \cite{seib,reviews}. 
Most of the results were first guessed at within field theory
and then checked to satisfy many non-trivial consistency
conditions. However, the organizing principle behind these dualities has
always been somewhat mysterious from the field theoretic perspective.
  
Recently there has arisen a fascinating connection between
supersymmetric gauge dynamics and string theory brane
dynamics  \cite{branereviews} which has the potential
for unifying our understanding of these dualities [4-13]. 
This stems from our ability to set up configurations of branes 
in string theory with supersymmetric gauge field theories living on the
world-volumes of branes in the low-energy limit.
The moduli spaces of
the gauge field theories are thereby encoded geometrically in the brane
set-up. Furthermore, in several cases it has been shown that the 
magnetic dual of a field theory can be obtained as the low-energy limit
of a brane configuration obtained by a continuous deformation of the
brane configuration corresponding to the electric theory. This
constitutes a derivation of field theoretic duality from string theory if the
infrared limit is unaffected by the deformation. An alternative approach
that allows one to derive non-trivial field theory results including
$N=1$ dualities has been developed using F-theory [14-21].

The strong coupling dynamics of the
low-energy limit of supersymmetric gauge theories can also be studied
using M-theory. In this approach the Dirichlet 4-branes (D4-branes) appearing
in Type IIA string theory constructions of field theories, are replaced by 
M-theory 5-branes wrapped around the compact eleventh dimension, while
the Neveu-Schwarz (NS) 5-brane of Type IIA string theory remains a
5-brane of M-theory. It has been  shown that four-dimensional $N=2$
supersymmetric gauge dynamics can then be represented in M-theory by a
single 5-brane surface, and  
 this surface is directly related to the
curves that appear in the solutions to the Coulomb branch
of the field theory \cite{klemm} \cite{witten1}. 
Subsequent developments for $N=2$ appear in refs. [24 -- 26]. 
The M-theory approach has also been generalized to
the study of moduli space in $N=1$ supersymmetric gauge
theories [27,28]. In the context of pure $N=1$ supersymmetric
Yang-Mills theory, Witten has  shown how the low-energy effective
superpotential can be computed from M-theory \cite{witten2}.
This was subsequently generalized to the case including matter
in ref. \cite{nos}. Witten also derived a hadronic
string from a special limit of M-theory, and this has also been
generalized to the case with matter \cite{hsz}. 
  
In this paper we derive Seiberg's dualities for four-dimensional $N=1$
supersymmetric QCD (SQCD) from M-theory. A central result is that
both the SQCD electric
theory and its magnetic dual (when there is one) are described by the
same M-theory configuration. Specifically, one can start from the string
theory brane configuration whose low-energy limit is the electric
theory, make the string theory coupling large enough to pass into
M-theory, change the parameter in the M-theory configuration which
corresponds to the (electric) strong interaction scale from very small
to very large (compared to the string scale), and then make the string 
coupling weak again. The result is the string theory brane configuration
whose low-energy limit is the magnetic dual field theory!  
 The advantage of our M-theory derivation of duality over the corresponding
string theory derivations is that the presence of the compact eleventh
dimension of M-theory smoothes the singular
situations that can occur as intermediate steps in the deformation of
the electric configuration to the magnetic configuration in string theory.
By going to M-theory we derive a
pleasing picture of how  D4-branes are created 
when NS 5-branes are passed through each other.  

The organization of this paper is as follows.
Section 2 reviews those features of SQCD which we use in the rest of the
paper, and sets some notation. Section 3 reviews the string theory
brane configuration whose low-energy limit is SQCD. Section 4 describes
the translation of this set-up to M-theory, largely following ref. [28].
 Section 5 shows how SQCD
duality emerges from the M-theory description, with a minor technical 
flaw  associated with our use of semi-infinite D4-branes.
This flaw is corrected in section 6, by using only finite D4-branes, in
the special case of equal quark masses. An interesting
associated subtlety which is resolved by non-perturbative bending of
D4-branes is discussed. Section 7 provides our conclusions.

\section{Review of SQCD}

Detailed derivations of the results quoted here can be found in
refs. [1, 2].
Consider an $N=1$ SQCD theory with $SU(N)$ gauge group and F flavors of
quarks ($Q_+$) and anti-quarks ($Q_-$). 
The standard (one-loop) formula for the strong interaction
scale is given by 
\begin{equation}
\Lambda^{3 N - F} = \mu^{3N-F} e^{ - (8 \pi^2/g_{SQCD}^2(\mu) + i \theta)},
\eqn{scale}
\end{equation}
where the phase is given by the CP-violating $\theta$ angle. 

The quark mass matrix is denoted
by $m$, and (for technical string/M-theory reasons) we will restrict our
attention to the case where the mass eigenvalues, $m_1, ... , m_F$ are
all non-vanishing. They can however be chosen to be arbitrarily small
and can then be usefully thought of as sources for quark bilinears in 
massless SCQD. If however, we take a quark mass to be very large, say
$m_F$, we can integrate out the massive quark and get a lower-energy
effective theory with
$F-1$ flavors and a strong scale given by 
\begin{equation}
\Lambda_L^{3N - F +1} = m_F \Lambda^{3N-F}.
\eqn{matching}
\end{equation}
The masses lift all the classical flat directions
of massless SQCD, which are parameterized by meson, (and for $F \geq N$)
baryon and
anti-baryon chiral superfields. As a consequence,  the baryon and
anti-baryon vevs are fixed at zero, while the vev of the meson fields 
defined by the gauge-invariant bilinear, 
\begin{equation}
M \equiv Q_+ Q_-,
\end{equation}
is diagonalized in the same basis as $m$, with eigenvalues,
\begin{equation}
\langle M \rangle_i = \frac{({\rm det} ~m)^{1/N} 
\Lambda^{(3N-F)/N}}{m_i}.
\eqn{Mvev}
\end{equation}

For $F < N$, the $\langle M \rangle_i$ exhibit runaway behavior in the
massless SQCD limit.
For $F = N$, we can approach the massless limit without
the vacuum running away. 
The $\langle M \rangle_i$
can be chosen arbitrarily by choosing the ratios $m_i/m_j$ as $m_i
\rightarrow 0$, subject only to the constraint 
\begin{equation}
{\rm det} \langle  M \rangle \equiv \prod_i \langle M \rangle_i = \Lambda^{2N}.
\end{equation}
This is just the quantum deformed moduli constraint that emerges when
approaching massless SQCD from massive SQCD, since, as mentioned above, 
the baryon vevs which
usually appear in the constraint are always at zero vev for arbitrary
but non-zero quark masses. 
For $F > N$  we can arrange arbitrary vevs for the mesons as
we approach the massless limit,
subject  only to the rank constraint, namely that only $N$ of the 
$\langle M \rangle_i$'s can be non-zero. 

The results described above are deduced quite concretely. On the other
hand  duality has to be guessed,  and then shown to satisfy a number of
consistency conditions. For $F > N+1$ the infrared behavior of 
SQCD is believed to have
a dual description \cite{seib} in terms of an $SU(\tilde{N})$ gauge group, 
$\tilde{N} = F-N$, with $F$ flavors of quarks and anti-quarks, and with
gauge-singlet mesons which are interpolated by the meson operators discussed
above. The meson fields of the dual theory are usually denoted by 
their interpolating operator, the difference in dimension being
compensated by a scale $\mu$. The dual effective theory has a superpotential
given by,
\begin{equation}
W_{dual} = {\rm tr}~ mM + \frac{1}{\mu} \tilde{Q}_+ M \tilde{Q}_-, 
\eqn{dualpot}
\end{equation}
at energies far above the $\langle M \rangle_i/\mu$. 
The strong interaction scale of the dual theory is given by 
\begin{equation}
\tilde{\Lambda}^{3 \tilde{N} - F} \Lambda^{3N - F} = (-1)^{\tilde{N}}
\mu^{F}.
\eqn{laladual}
\end{equation}
The scale $\mu$ appearing in \eq{dualpot} and \eq{laladual} is arbitrary
in the sense that it does not affect the low-energy regime
where duality is expected to hold. Below
the dual quark masses  $\langle M \rangle_i/\mu$, the dual quarks 
 can be integrated out of the theory and the dual gauge
theory undergoes gaugino condensation, resulting in,
\begin{equation}
W_{dual} = {\rm tr}~ mM +  
\tilde{N} \tilde{\Lambda}^{(3 \tilde{N}-F)/\tilde{N}} ({\rm det}
\frac{M}{\mu})^{1/\tilde{N}}.  
\end{equation}
Note that this  is independent of $\mu$ when expressed in terms of
$\Lambda$ using \eq{laladual}. 
The case $F= N+1$ is a somewhat degenerate case of duality with the
trivial dual gauge group ``$SU(1)$'' whose ``dual quarks'' are just
baryons. 

The quark masses break all non-abelian chiral
symmetries. We will focus our attention on two $U(1)_R$-symmetries, differing
only in the charges assigned to the matter multiplets:

\beq
 \begin{array}{c|cccc}
    & Q_+, Q_- & M & m & \La^{3N-F} \\[.1in] \hline
&&&&\\[-.1in]
  R_v & 0 & 0 & 2 & 2(N-F) \\
  R_ w& 1 & 2  & 0  & 2N 
\end{array}
\eqn{SQCDtable}
\eeq

Both symmetries are anomalous, symmetry transformations result in
shifts of the $\theta$ angle. These shifts are indicated by assigning
spurious charges to the strong scale $\Lambda$ (see
\eq{scale}). Periodicity of physics in $\theta$ then implies that the anomaly
breaks the $R_v$-symmetry to its ${\bf Z}_{2(N-F)}$ subgroup, while the
$R_w$-symmetry is broken down to its ${\bf Z}_{2N}$ subgroup. 
The $R_v$-symmetry is also explicitly broken by $m$, so that
$R_v$-transformations induce rotations of $m$. This is indicated by the
spurious charge assigned to $m$. The ${\bf Z}_{2N}$ symmetry, though an
exact dynamical symmetry, is spontaneously broken down to ${\bf Z}_2$ by
gaugino condensation. 

In this paper we shall consider $F < 3N$ so that SQCD is asymptotically
free and has interesting non-perturbative
effects. 

\section{ The Type-IIA string theory set-up}

From now on we shall work in units in which the string mass scale is set
to one, $m_s \equiv 1$. 
Our SQCD theory can be described as the low-energy
limit of D4-branes suspended between Neveu-Schwarz 
(NS) 5-branes in Type IIA string
theory. The ten-dimensional  
configuration of branes we will consider is depicted in
Fig. 1. It is similar to the configurations with D6-branes used to
study SQCD in ref. \cite{egk} but employs semi-infinite fourbranes
as suggested in \cite{witten1} [28].
All the branes occupy the four dimensions spanned
by the $x^0, x^1, x^2, x^3$ directions and all of them sit at $x^9 =
0$, so these five dimensions have been suppressed in the figure. The
$x^4$ and $x^5$ coordinates are conveniently paired into a complex
coordinate, as are the $x^7$ and $x^8$ coordinates,
\begin{equation}
v = x^4 + i x^5, ~~ w = x^7 + i x^8.
\end{equation}
These complex dimensions are represented schematically as real
dimensions in Fig. 1. (One can consider the imaginary components of $v$
and $w$ to have been suppressed.) 
The NS $5_v$-brane sits at $w=0$ and fills the  $v$-plane,
while the NS $5_w$-brane sits at $v=0$ and fills the $w$-plane. The
two 5-branes are separated in the $x^6$-direction by a distance
$S_0$. $N$ coincident D4-branes are suspended between the 5-branes at 
$v = w = 0$. $F$ semi-infinite D4-branes at $w = 0$ and $v = v_1, ... ,
v_F$ come in from $x_6 = - \infty$ and attach to the NS 5-brane.

\begin{figure}[t]
\centerline{\epsfxsize=4 in \epsfbox{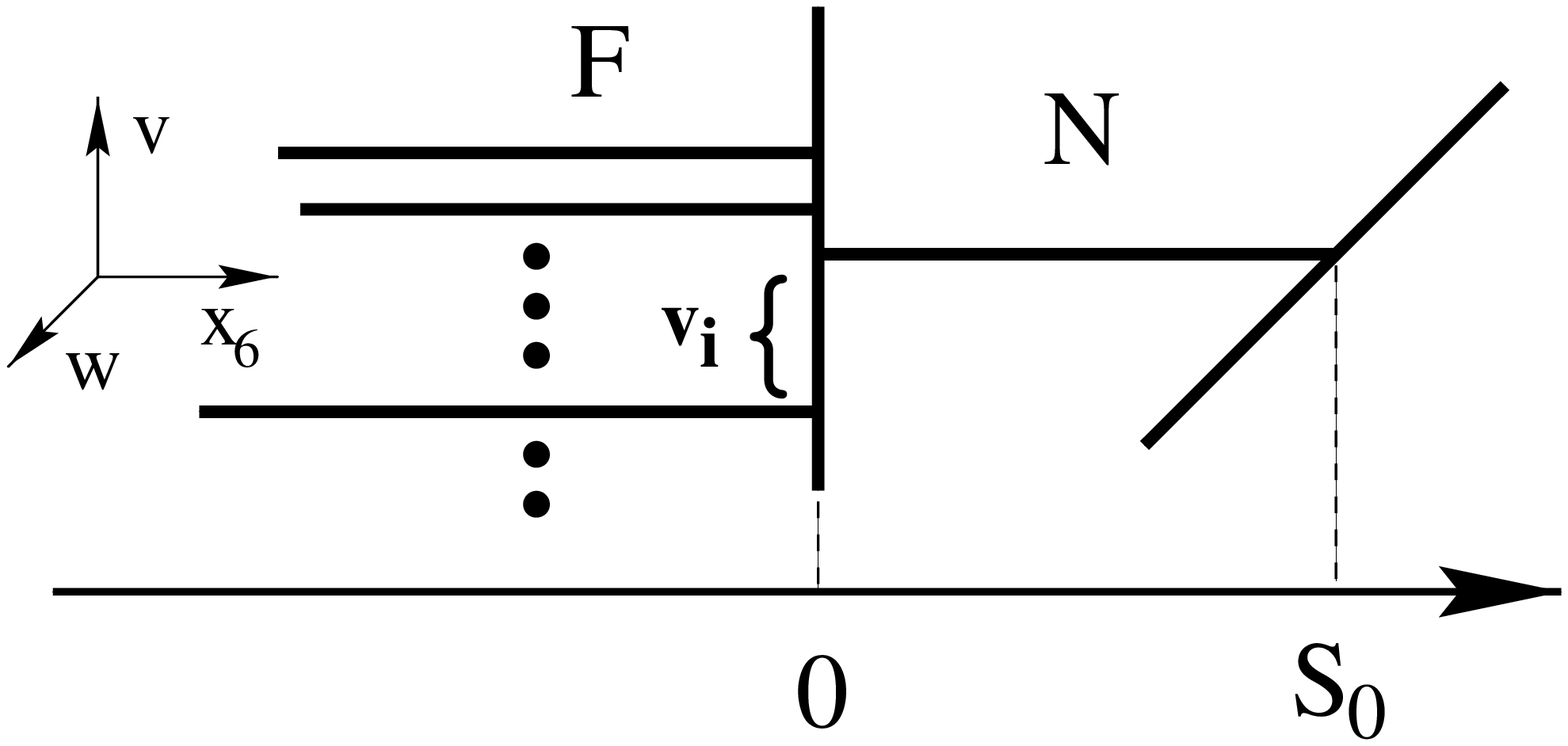}}
\vskip.2in
\noindent
Fig 1. {\it The classical string theory brane configuration corresponding to
SQCD. The positions $v_i$ of the semi-infinite D4 branes determine
the bare masses of the quark fields.}
\vskip .2in
\end{figure}

The low-energy particle content essentially consists of the (nearly)
massless strings with ends attached to D4-branes. There are $N^2$
massless vector fields corresponding to the short strings connecting one
of the $N$ finite D4-branes with another. The five-dimensional strip
occupied by  these states is effectively
four-dimensional if $S_0$ is taken very small, and this is identified with
the four-dimensional spacetime of the low-energy limit. The massless
vector fields fields form the gauge fields of a $U(N)$ gauge
symmetry. The $U(1)$ factor corresponding to the trace of
the $U(N)$ group has been argued to be frozen by an infrared
singularity on the NS5 branes \cite{witten1}.
However, as has been argued \cite{egkrs}, the $U(1)$ can be
resurrected by adding a D4-brane
at infinity. Even if it is present it is infrared free and
decouples from the $SU(N)$ dynamics.
We shall ignore the possible $U(1)$ factor
in the following. The SQCD gauge coupling is given by 
\begin{equation}
\frac{8 \pi^2}{g_{SQCD}^2} = \frac{S_0}{g_s}.
\end{equation}
Strings can also connect the $F$ semi-infinite D4-branes and the $N$
finite D4-branes. For small $v_i$ this results in
 $F$ flavors of quark and anti-quark multiplets, with
mass parameters given by 
\begin{equation}
m_i = v_i.
\end{equation}
 The price of this simple set-up is that all
the masses must be taken non-zero. Massless SQCD requires the
introduction of more branes.  
The matter multiplets live in the effectively four-dimensional spacetime
as well
because they correspond to short strings with one end on the finite
D4-branes.
 There are also 
 short strings connecting the semi-infinite D4-branes with each
 other. These correspond to unwanted light states living in {\it five}
 (semi-)infinite dimensions! These  states do couple to the quark fields
 living on the four-dimensional boundary of their five-dimensional
 world, but these couplings are non-renormalizeable (suppressed by
 powers of $m_s$) and one can hope that their infrared effects do not
 interfere with the four-dimensional SQCD dynamics. 

\section{The M-theory set-up}

In the above set-up, the non-perturbative SQCD effects are also
non-perturbative in string theory. To get at these  
effects we go to M-theory, largely following ref. [28]. This is the regime
where we take $g_s$ to be large,  opening up a new
compact $x^{10}$ dimension with radius,
\begin{equation}
R = g_s.
\end{equation}
Now, as long as $g_s$ is perturbatively weak, variations in it do not
affect the low-energy limit of the theory, which remains SQCD. Even
$g_{SQCD}$ can be kept fixed by varying $S_0$ with $g_s$. It is assumed
that the very low energy limit is unaffected by taking $g_s$ large
enough to enter semi-classical M-theory. We can then use M-theory to
solve for the non-perturbative long-distance dynamics of SQCD.

In M-theory, the brane configuration of Fig. 1 becomes ``thickened'' in
the new dimension, so that D4-branes also become 5-branes, but wrapped around
the $x^{10}$-direction. They smoothly connect to the pre-existing 5-branes.
The resulting smooth five-dimensional surface has the form ${\bf R}^4
\times \Sigma$, where ${\bf R}^4$ is a copy of four-dimensional spacetime
and $\Sigma$ is a complex one-dimensional curve  in a complex
three-dimensional space spanned by $v, w$ and $t \equiv e^{-s/R}$,
where,
\begin{equation}
s \equiv x^6 + i x^{10}.
\end{equation} 
We can determine the M-theory curve corresponding to Fig. 1 by noting
that if the $5_w$-brane were removed we  would have semi-infinite
branes on either side of the $5_v$-brane, which is a degenerate case of
the $N=2$ supersymmetric configurations dealt with in \cite{witten1}. 
The curve for
this case has the form,
\begin{equation}
t~ \prod_{i = 1}^F (v - v_i)  - \xi v^N = 0, ~~ w = 0.
\eqn{vt}
\end{equation}
Now let us include the presence of the $5_w$-brane. In M-theory this
corresponds to points of the curve $\Sigma$ with large $w$ and small
$v$, that is $w$ has a simple pole in $v$. Thus,
\begin{equation}
v w = \zeta.
\eqn{vw}
\end{equation}

\Eq{vt} and \eq{vw} define the M-theory curve corresponding to Fig. 1, in the
sense that for small $R$ ($g_s$),  the curve
degenerates to Fig. 1. This is illustrated in Fig. 2, where the curve is
plotted in the plane of $x^6$ and (the real part of) $v$ for specific
choices of $\xi, \zeta$ and small $R$. M-theory has smoothed out the
singular brane junctions of string theory in a manner which is
non-perturbative in string theory. This smoothing out also
contains non-perturbative information about the low-energy limit of SQCD. 

\begin{figure}[t]
\centerline{\epsfxsize=4 in \epsfbox{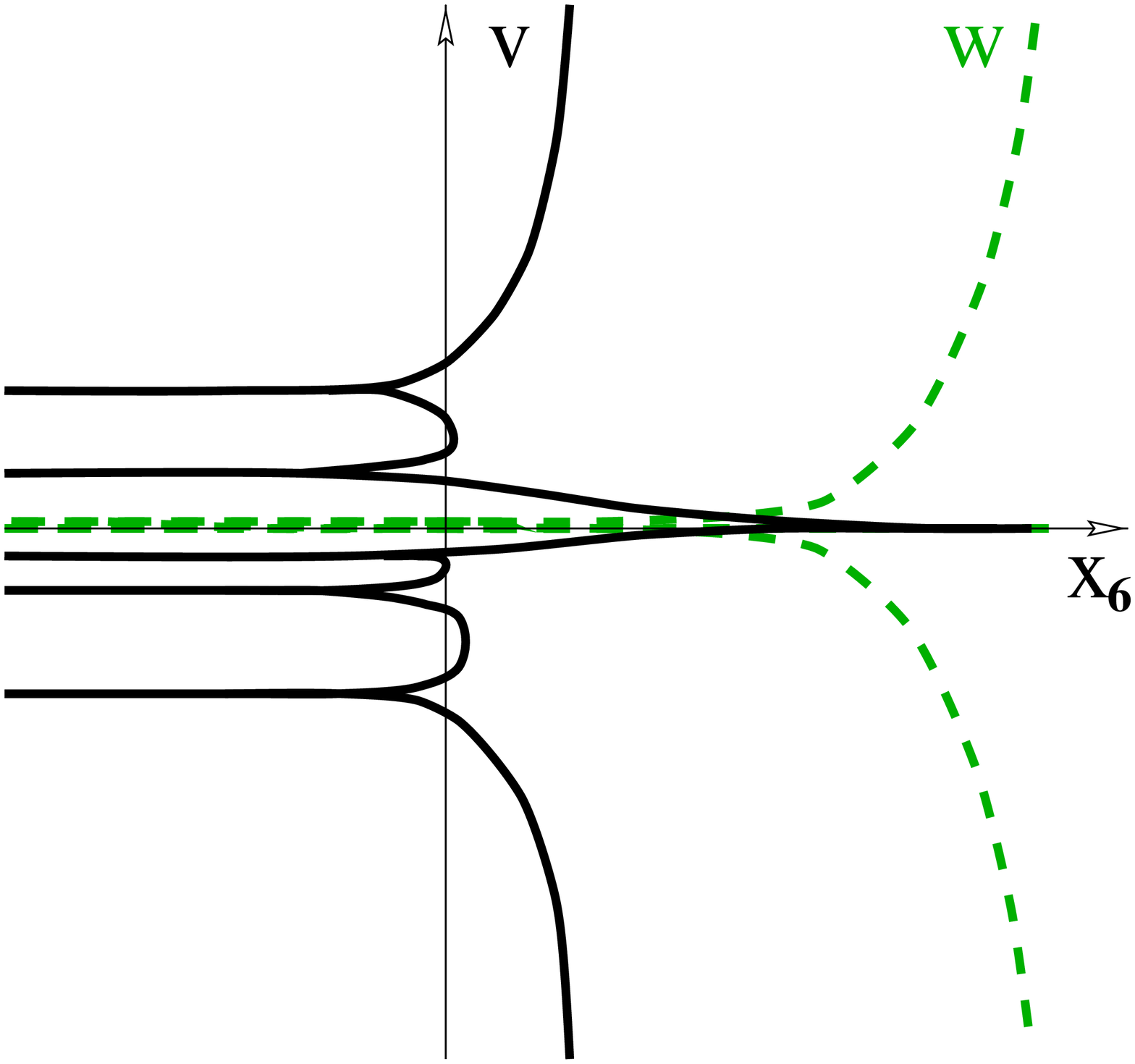}}
\vskip.2in
\noindent
Fig 2. {\it M theory ``thickened'' 5-brane configuration which
reduces to Fig. 1. in the string theory limit.}
\vskip .2in
\end{figure}

The parameters of the M-theory curve 
can be identified in terms of SQCD parameters as
follows \cite{biksy}:
\begin{equation}
\xi =  (\prod_i m_i)^{(F-N)/F}, ~~ \zeta =  \Lambda^{(3N -F)/N} 
(\prod_i m_i)^{1/N}, ~~ v_i = m_i \ .
\eqn{parameterid}
\end{equation}
There are some  $g_s$-dependent coefficients on the
right-hand sides which we have omitted because they are unimportant for
our story. The curve then becomes,
\begin{eqnarray}
t ~\prod_i (v -  m_i)  &-& (\prod_i m_i)^{(F-N)/F} v^N = 0, \nonumber \\
v w &=&  (\prod_{i} m_i)^{\frac{1}{N}} \Lambda^{(3N -F)/N}.
\eqn{thecurve}
\end{eqnarray}
Rotations of the $v$ and $w$
planes are identified with the $R$-symmetries discussed in section 2, by
assigning $R_v$-charge two to $v$ and zero to $w$, and assigning
$R_w$-charge two to $w$ and zero to $v$. 

Now, the symmetries of the
asymptotic behavior of the M-theory curve for large $v$ or $w$ 
correspond to the symmetries
of perturbative string theory and SQCD, while the symmetries of the
whole M-theory curve  correspond to the symmetries of the
non-perturbative vacuum. With this in mind one can see the neat fit of
the parameter identification above. The asymptotic behavior of the curve
for large $v$ is given by
\begin{equation}
v^{F-N} t - (\prod_i m_i)^{(F-N)/F} \sim 0, ~~ w \sim 0,
\eqn{largev}
\end{equation}
which is symmetric under $R_w$ and $R_v$ when $m$ is taken to transform
spuriously as in \eq{SQCDtable}. 
For large $w$ the curve becomes,
\begin{equation}
w^N t - (-1)^F (\prod_i m_i)^{(F-N)/F} \Lambda^{3F-N} \sim 0, ~~
v \sim 0, 
\eqn{largew}
\end{equation} 
which is not generically $R_{v,w}$-symmetric when $m$ transforms
spuriously. Instead the spurious rotations of $\Lambda$ given by
\eq{SQCDtable} result.
However we see that an $R_v$-rotation by angle $\pi/(N-F)$
does leave the curve invariant (when $m$ transforms spuriously), while
an $R_w$-rotation by angle $\pi/N$ also leaves the curve invariant,
corresponding to the non-anomalous ${\bf Z}_{2(N-F)}$ and ${\bf Z}_{2N}$
symmetries respectively. The exact curve, eq. (4.6), however, does not
respect these discrete symmetries, thus reproducing the physics of
gaugino condensation. 

One can also consider taking a quark mass, $m_F$, to be very large for a
fixed but arbitrary region of $v$ and $w$. 
It is straightforward to see
that \eq{thecurve} then reduces to,
\begin{eqnarray}
\kappa t \prod_i^{F-1} (v -  m_i)  &-& (\prod_i^{F-1} m_i)^{(F - 1 -N)/(F-1)}
v^N = 0, \nonumber \\
\kappa  &=& - m_F^{N/F} (\prod_i^{F-1} m_i)^{-N/[F(F-1)]},
\nonumber \\
v w &=&  (\prod_{i}^{F-1} m_i)^{\frac{1}{N}} \Lambda_L^{(3N -F+1)/N},
\end{eqnarray}
where $\Lambda_L$ is given by \eq{matching}. $\kappa$ does not transform
spuriously under the $R$-symmetries and can be absorbed into $t$ by a
trivial shift of the $s$-coordinate origin. We thereby arrive at the
curve corresponding to the effective field theory with the massive quark
field integrated out. This check, and the check of $R$-symmetries above,
uniquely specify the parameter identification of \eq{parameterid}.

Having identified the parameters of our M-theory curve, we can express
the separation of NS 5-branes in terms of these parameters by 
studying the $R \rightarrow 0$ limit. Comparing the large $v$ behavior
of the curve, \eq{largev}, which corresponds to the NS $5_v$-brane,
with the large $w$ behavior, \eq{largew},
which corresponds to the NS $5_w$-brane, we see that
the relative separation of these $5$-branes in the $s$-directions,
$S_0$, satisfies 
\begin{equation}
e^{- S_0/R} = \Lambda^{3N-F}. 
\eqn{lambdaands0}
\end{equation}
$S_0/R$ is now generalized from its usage in section 3 to be complex, 
 the imaginary part being an angular separation in the
 $x^{10}$-direction. By eq. (2.1), 
\begin{equation}
\frac{8 \pi^2}{g_{SQCD}^2} + i \theta = \frac{S_0}{R}, 
\end{equation} 
where the gauge coupling is renormalized at the string scale multiplied by
a function of $g_s$. This reproduces the string theory result eq. (3.2),
once we substitute eq. (4.1). 
At loop-level in SQCD (and string theory) the gauge coupling
runs. Witten has pointed out \cite{witten1} that this feature is reflected in
M-theory by the asymptotic logarithmic bending of the curve for large
$v$ and large $w$. This bending can be seen from eq. (4.7) and (4.8). As $R
\rightarrow 0$ this bending is reduced, and flat 5-branes such as those
depicted in Fig. 1 emerge in the limit if $S_0$  is kept fixed. This is
reasonable since it corresponds to taking $g_{SQCD} \rightarrow 0$ where
the $\beta$-function vanishes.

\section{Duality from M-theory}

We begin by noting that the M-theory curve encodes the 
non-perturbative vacuum expectation values of the meson operators,
\eq{Mvev}, in a rather suggestive way. 
(Recall that the baryon and anti-baryon vevs are zero for
non-zero quark masses). The SQCD $M$ eigenvalues emerge upon solving
for the general $t-w$ relationship
in the curve, \eq{thecurve} \cite{biksy}. 
\begin{eqnarray}
(-1)^F t \prod_i (w - \langle M \rangle_i) &-& 
(\prod_i \langle M \rangle_i)^{(F- \tilde{N})/F} w^{\tilde{N}} = 0, 
\eqn{tw}
\end{eqnarray}
where $\tilde{N} = F-N$. As explained in section 2, as $m \rightarrow 0$ 
the $\langle M \rangle_i$ display runaway behavior for $F < N$, a
quantum deformed moduli space for $F = N$, and the rank constraint on
the moduli for $F > N$. 

Once the factor of $(-1)^F$ is absorbed into $t$  by a trivial
redifinition of the origin of the $s$-coordinate, \eq{tw} looks just
like part of a curve for a gauge group $SU(\tilde{N})$ with $F$
flavors. Indeed the $v-w$ relationship in \eq{thecurve} also obeys this
duality as can be seen by rewriting it in terms of $\langle M \rangle_i$
and $\tilde{\Lambda}$ (as given by \eq{laladual} for  $\mu = 1$), 
\begin{equation}
v w = - (\prod_{i} \langle M \rangle_i)^{1/\tilde{N}} 
\tilde{\Lambda}^{(3 \tilde{N} -F)/\tilde{N}}.
\eqn{vwdual}
\end{equation}
\Eq{tw} and \eq{vwdual} are just a rewriting of our old
curve, \eq{thecurve}, but we see
that it is also the curve corresponding to the dual gauge group and dual
quarks with masses $\langle M \rangle_i$.
Of course this is only true if $\tilde{N} > 0$, with $\tilde{N} =
1$ being the degenerate case of duality mentioned in section 2. 
This M-theoretic duality is the central observation of this paper. 

We have found that one and the same
 curve describes the M-theory  configuration associated
with the string theory set-up of Fig. 1 as well as the string theory
set-up of Fig. 3. We have already illustrated in Fig. 2 the sense in
which the curve degenerates to Fig. 1 as $R$ becomes small. Clearly it
cannot simultaneously degenerate to Fig. 3. To understand this puzzle
note that in order to degenerate to Fig. 1 as $R \rightarrow 0$ we need to
keep $m_i$ and $S_0$ fixed. By \eq{lambdaands0} this means
that $\Lambda \rightarrow 0$ exponentially fast.
This makes physical sense because it implies that the quarks and
gluons are weakly coupled string modes at the string scale, and we
expect to see them as short strings in the classical approximation of
Fig. 1. However this is not the regime in which dual quarks and gluons
are weakly coupled at the string scale, so we should not expect the
classical setup of Fig. 3 to be valid. 
Fig. 3 is the classical string theory set-up
corresponding to having dual quarks and gluons being short strings
attached to D4-branes, weakly coupled at
the string scale. For $3N > F > 3N/2$ the dual theory is also
asymptotically free and so the dual quarks and gluons should be weakly
coupled at the string scale  if and only if $\tilde{\Lambda} \ll
1$, which corresponds to $\Lambda \gg 1$. For $3N/2 > F > N$ the dual
theory is infrared free and the dual quarks and gluons are weakly
coupled at the string scale if it lies far below the Landau pole,
$\tilde{\Lambda}$. Again this corresponds to taking $\Lambda \gg
1$ as can be seen from eq. (2.7). 
Therefore in both cases we expect to see the dual quarks and gluons
weakly coupled at the string scale by taking $\Lambda \gg 1$. Note
that this corresponds to $S_0 < 0$ by \eq{lambdaands0}, corresponding
to reversing the order of the two NS 5-branes. Indeed, if $S_0 < 0$ and
$\langle M \rangle_i$ are fixed as $R \rightarrow 0$, the curve
degenerates to Fig. 3, as illustrated in Fig. 4. 

\begin{figure}[t]
\centerline{\epsfxsize=4 in \epsfbox{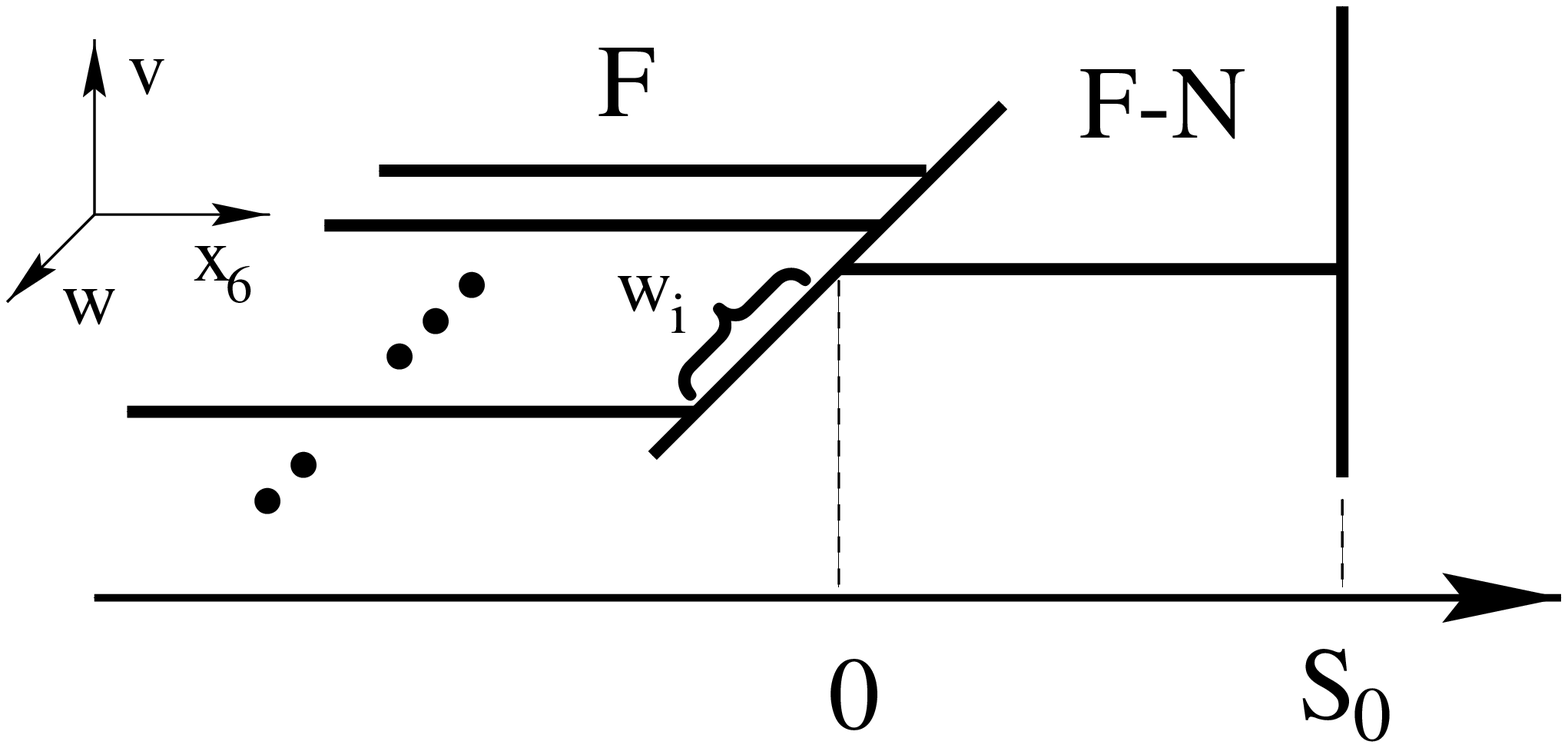}}
\vskip.2in
\noindent
Fig 3. {\it The classical string theory brane configuration corresponding to
the dual of SQCD. The positions $w_i$ of the semi-infinite D4 branes
determine the expectation values for the meson field.}
\vskip .2in
\end{figure}

\begin{figure}[t]
\centerline{\epsfxsize=4 in \epsfbox{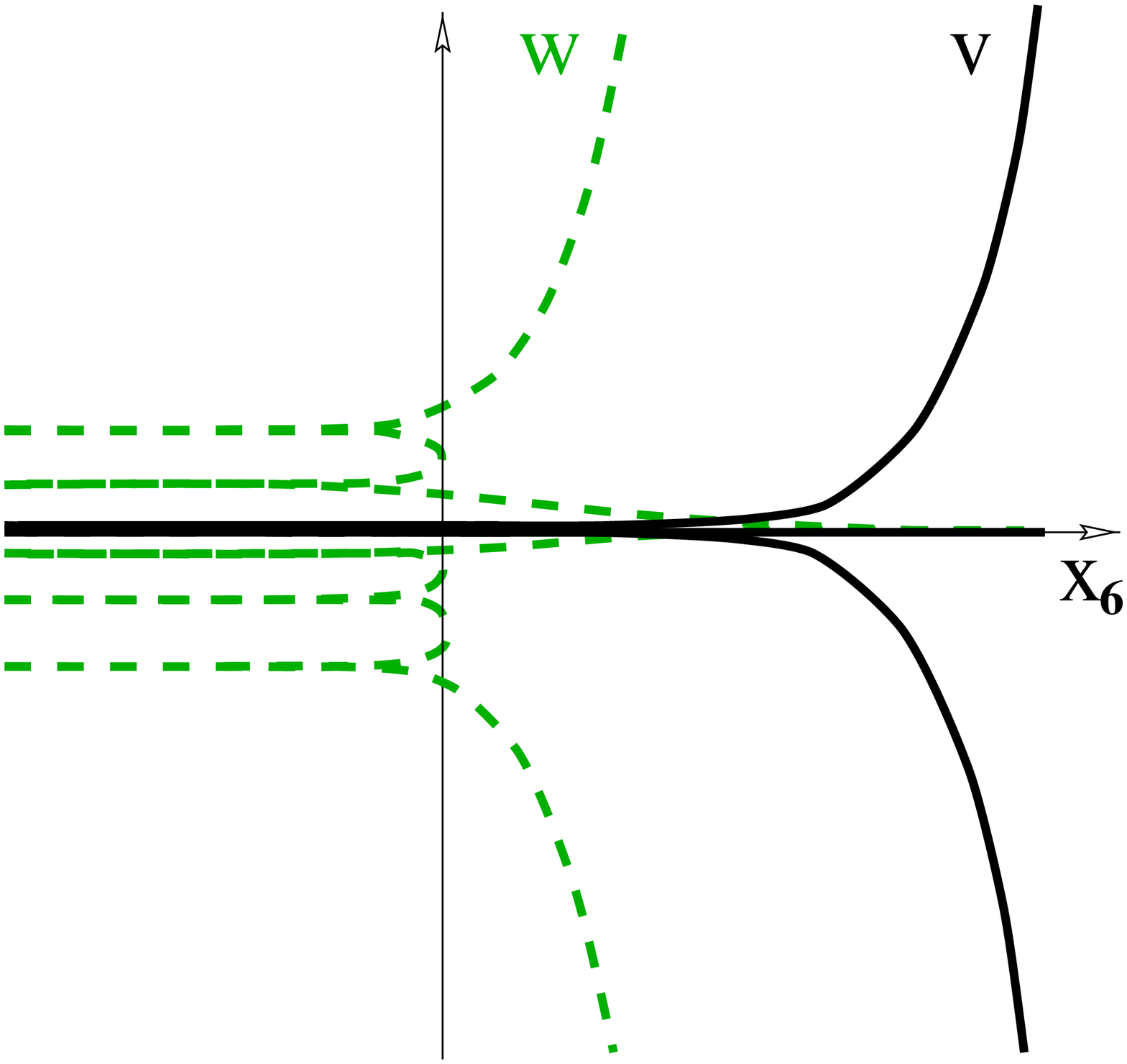}}
\vskip.2in
\noindent
Fig 4. {\it  M theory ``thickened'' 5-brane configuration which
reduces to Fig. 3. in the string theory limit.}
\vskip .2in
\end{figure}

The fact that we must take $\Lambda \gg 1$ to ``see'' duality has a
simple interpretatation in the case $3N/2 > F > N$. Here 
SQCD becomes strongly coupled in the infrared (if the meson vevs
are smaller than $\Lambda$). We can think of string/M-theory as
providing  a supersymmetric ultraviolet cutoff for our field
theory. The continuum limit corresponds then to $\Lambda \ll 1$. We
can imagine lowering our cutoff by integrating out higher energy
physics, until $\Lambda \gg 1$. By this stage we have induced a lot of
non-renormalizeable effects in the 
effective theory. Let us simply
discard these effects. We can hope that the qualitative strong
interaction physics is unchanged by this drastic truncation of the theory.
This leaves the gauge coupling which is now very large at the cutoff, so we
are justified in doing a strong coupling expansion. This is precisely
the nature of strong coupling expansions of (non-supersymmetric) QCD
formulated on the lattice. Much of the qualitative non-perturbative physics of
confinement is thereby reproduced, though quantitative details are
not. 
This is also the situation we find ourselves in. We have
$\Lambda \gg 1$, which justifies a strong coupling expansion. The
strong coupling expansion of the associated string theory is provided by
M-theory. Just as in lattice strong-coupling expansions, M-theory does
not provide a {\it quantitatively} accurate solution of all aspects of 
SQCD (for
example it will not accurately give the mass ratios of hadrons)
\cite{witten2}, but it
provides us with the correct infrared limit, in particular SQCD duality.

While we have already found that the meson vevs are contained in
M-theory, it is
disturbing that the low-energy limit of the dual set-up in
Fig. 3 does not appear to permit short strings that correspond to the
meson fluctuations. These light meson degrees of freedom are required in the
dual field theory. Light string modes with the right flavor quantum
numbers do appear connecting the semi-infinite branes, but as mentioned
in section 3, these modes live in five dimensions and
their coupling to the four-dimensional physics is very weak at low
energies. In fact we believe that the problem of the missing mesons is
an artifact of our use of semi-infinite branes to set up SQCD and the
delicate nature of the decoupling of the unwanted five-dimensional
fields. To get around this we must regulate the
semi-infinite branes somehow. This problem is discussed for the simple
case of equal quark masses in section 6, and indeed the requisite mesons
then naturally emerge. 

We have shown that by varying parameters of the M-theory curve we can
pass smoothly between Fig. 2 and Fig. 4. The analog of this in the
string theory limit is passing from Fig. 1 to Fig. 3. 
Apparently we have moved the 5-branes through
each other in this process and changed the number of D4-branes between
them. This is just the type of move that has been proposed within string
theory, in various contexts [3-12], for deriving field theory
dualities. Such moves in string theory involve certain singular
intermediate configurations, which in the present case corresponds to the
stage at which the two 5-branes  intersect.
Recalling \eq{lambdaands0} we see that the SQCD
coupling formally blows up there. By contrast in our M-theory
derivation of the move there is no singular intermediate
stage. In this sense M-theory has bought us a rather vivid resolution
of the singularities encountered in the string theory moves and a
simple derivation of the fact that D4-branes are created between
NS 5-branes as they are passed through each other. 

\section{Duality with finite D4-branes}

In order to avoid the pathologies associated with semi-infinite branes
and identify the missing meson of dual SQCD we now present a modified
brane setup with D4 branes which are of finite extent in the $x_6$
direction. With a finite $x_6$ direction we can expect to find a well defined
four-dimensional effective field theory on the D4-branes.

There are several choices for cutting off the semi-infinite branes. One
possibility which we are not going to explore in this paper is to have
them end on D6-branes which are located at large negative $x_6$ values.
Alternatively the D4-branes could
end on an NS5 brane at large negative $x_6$
as depicted schematically in Fig. 5.a. The problem with this setup is
that the D4-branes are now free to slide along the two parallel NS
5-branes.  This implies a moduli space of vacua parameterized by the positions
of the D4-branes with an associated chiral superfield
which transforms as an
adjoint of $SU(F)$ flavor. The existence of this scalar can also be seen by
noting that the theory on the D4-branes between the parallel
NS 5-branes is approximately $N=2$ supersymmetric thus having an
adjoint chiral multiplet in addition to the $N=1$ gauge multiplet.
Clearly this theory differs from our target theory SQCD. 

\begin{figure}[t]
\centerline{\epsfxsize=4 in \epsfbox{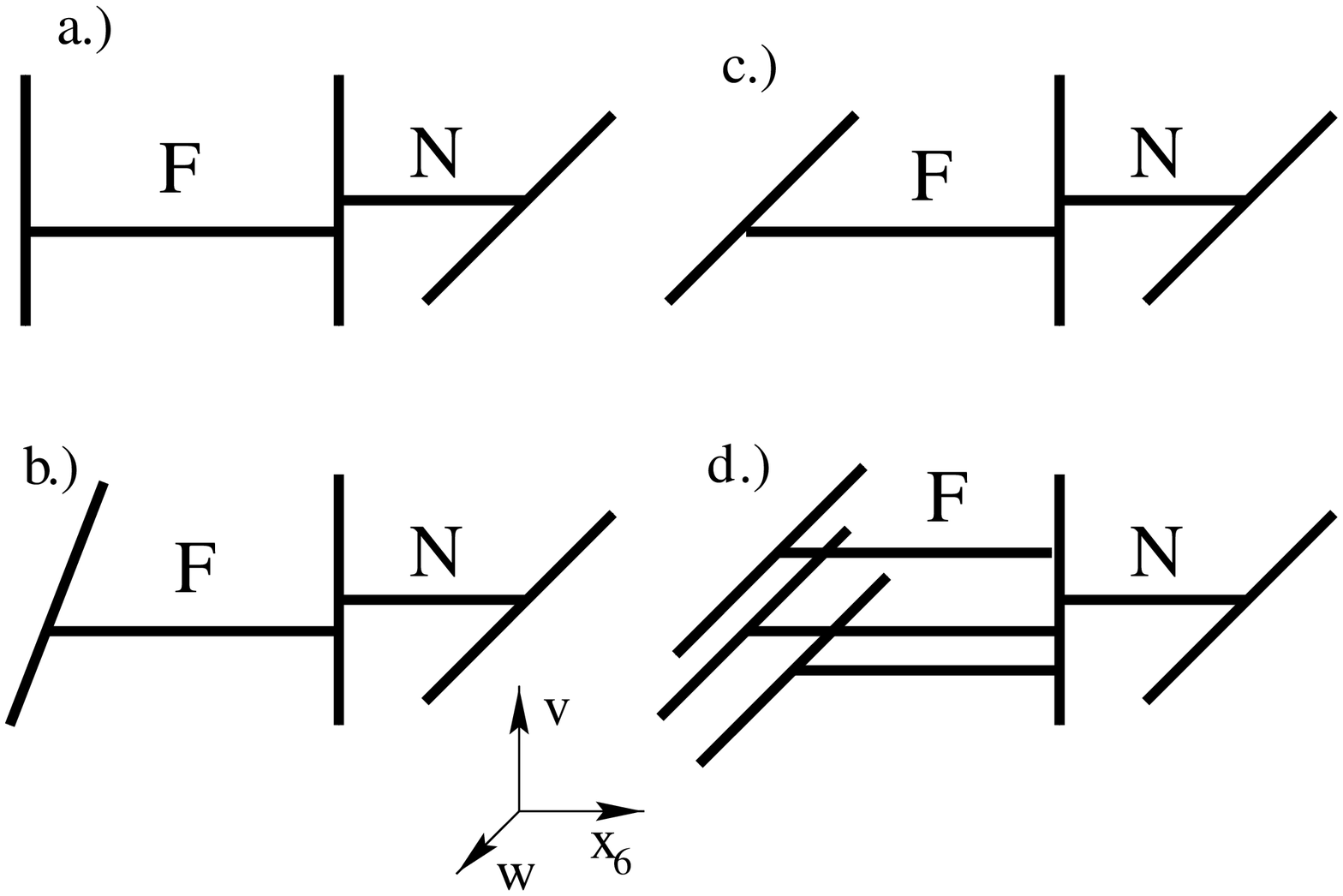}}
\vskip.2in
\noindent
Fig 5. {\it Different possibilities for terminating the flavor
branes at large negative $x_6$.}
\vskip .2in
\end{figure}

However, it is possible to give the adjoint superfield a mass by rotating
the NS 5-brane located at large negative $x_6$ into the $w$ direction
as shown in Fig. 5.b. For rotation angles between zero and $\pi/2$
this corresponds to a finite mass for the adjoint [8] \cite{hoo}.
This mass goes to infinity as the rotation angle
is taken to be $\pi/2$ (Fig. 5.c.), leaving us with an
effective $N=1$ gauge theory with no adjoint scalar on the
``flavor branes''. This is the setup that we want to focus on.
At length scales larger than the typical $x_6$ dimensions the low
energy effective theory is SQCD with $N$ colors and $F$ flavors with
a weakly gauged flavor group $SU(F)$. By taking the flavor
D4-branes longer than the color D4-branes we can make the scale
of the $SU(F)$ gauge group exponentially smaller than the scale of
the $SU(N)$.

Note that the setup of Fig. 5.c. only allows a common
mass for all quarks, but we could easily accomodate more general mass
terms by attaching an NS 5-brane to each of  the flavor branes
individually (Fig. 5.d.). To simplify the considerations, we
consider only the common mass case which has all the features
we wish to demonstrate. 

\subsection{The Field Theory}

Let us first investigate briefly the properties of  the low energy field
theory using field theory techniques. The field content and (spurious)
symmetries are
\beq
 \begin{array}{c|cccc}
    & SU(N) & SU(F) & R_v & R_w \\[.1in] \hline
&&&&\\[-.1in]
  Q_+, Q_- & \Yfund\ ,\ \Ybarfund & \Yfund\ ,\ \Ybarfund  & 0 & 1 \\
  m    & 1 & 1  & 2  & 0 \\
  \LaN^{3N-F}  & 1 & 1 & 2(N-F) & 2N \\
  \LaF^{3F-N}  & 1 & 1 & 2(F-N) & 2F 
\end{array}
\eqn{NFtable}
\eeq
with a tree level superpotential $W=m \tr Q_+Q_-$.
The theory has a discrete set of supersymmetric vacua with expectation
values for the meson 
\beq
\tr \langle M \rangle 
=F m^{(F-N)/N}\LaN^{(3N-F)/N} + N m^{(N-F)/F}\LaF^{(3F-N)/F}, 
\eqn{FNvev}
\eeq
which can be calculated from gaugino condensation and scale
matching relations \eq{matching}. Note that by choosing
$\LaF \ll (\LaN, m)$ one can decouple the contribution of the dual
gauge group and reproduce the SQCD meson vev \eq{Mvev}
for the case of a common mass. For $F > N$ the theory has a
dual description with an $SU(F-N)$ dual gauge group and $F$ flavors
transforming under the weakly gauged $SU(F)$.
The dual also has a fundamental meson field which transforms
as an adjoint, $M_{\rm adj}$, plus a singlet, 
$\tr M$, under $SU(F)$ which are both  coupled to the
dual quarks in the superpotential 

\beq
W= m\ \tr M + \frac{1}{\mu} [\frac{1}{F} \tr M\  (\twi Q_+ \twi Q_-) + 
\twi Q_+ M_{\rm adj} \twi Q_-] \ .
\eeq

Classically, this dual theory is an O'Raifeartaigh model
which breaks supersymmetry. This
is most easily seen by noting that the composite matrix
$\twi Q_+ \twi Q_-$ with the $SU(F-N)$ color indices
contracted cannot have an expectation value of rank
greater than $F-N$ because the rectangular matrices $\twi Q_+$
and $\twi Q_-$ are of rank less or equal to $F-N$. But the
equations of motion for $M_{\rm adj}$ and tr$M$ require 
$\twi Q_+ \twi Q_-$ to have rank $F$.
Quantum mechanically, supersymmetry is restored by
nonperturbative $SU(F-N)$ dynamics which generates the
superpotential
\beq
W_{\rm dyn}=\twi N\det (\frac{M_{\rm adj}+\tr M/F}{\mu})^{1/\twi N} 
\LaFN^{(3 \twi N -F)/\twi N},
\eqn{wdynFN}
\eeq
where $\LaFN$ is the scale of the dual gauge group. Minimizing
the potential including this term reproduces the supersymmetric vacua
of the electric theory. This resurrection of supersymmetry
by quantum effects in the dual has a nice
interpretation in the string/M-theory picture which will be discussed in
the next subsection.

Note that in this discussion we have ignored the
``weakly'' gauged $SU(F)$ symmetry. The justification for this is
not completely obvious, and requires a careful consideration of the
scales of the problem.
For example, for $F \ge 3/2 N$ the $SU(F-N)$ gauge coupling
and the Yukawa couplings (with $\LaF \ll m \ll \LaN$) are expected
to flow to an approximate interacting fixed point (above $m$). 
However when scaling to even lower energies
the $SU(F)$ gauge coupling increases and the theory moves away
from the SQCD fixed point. Thus our theory is only a good
approximation to SQCD and its dual if we look at energies large compared
to $\LaF$.

\subsection{M-theory}

\begin{figure}[t]
\centerline{\epsfxsize=4 in \epsfbox{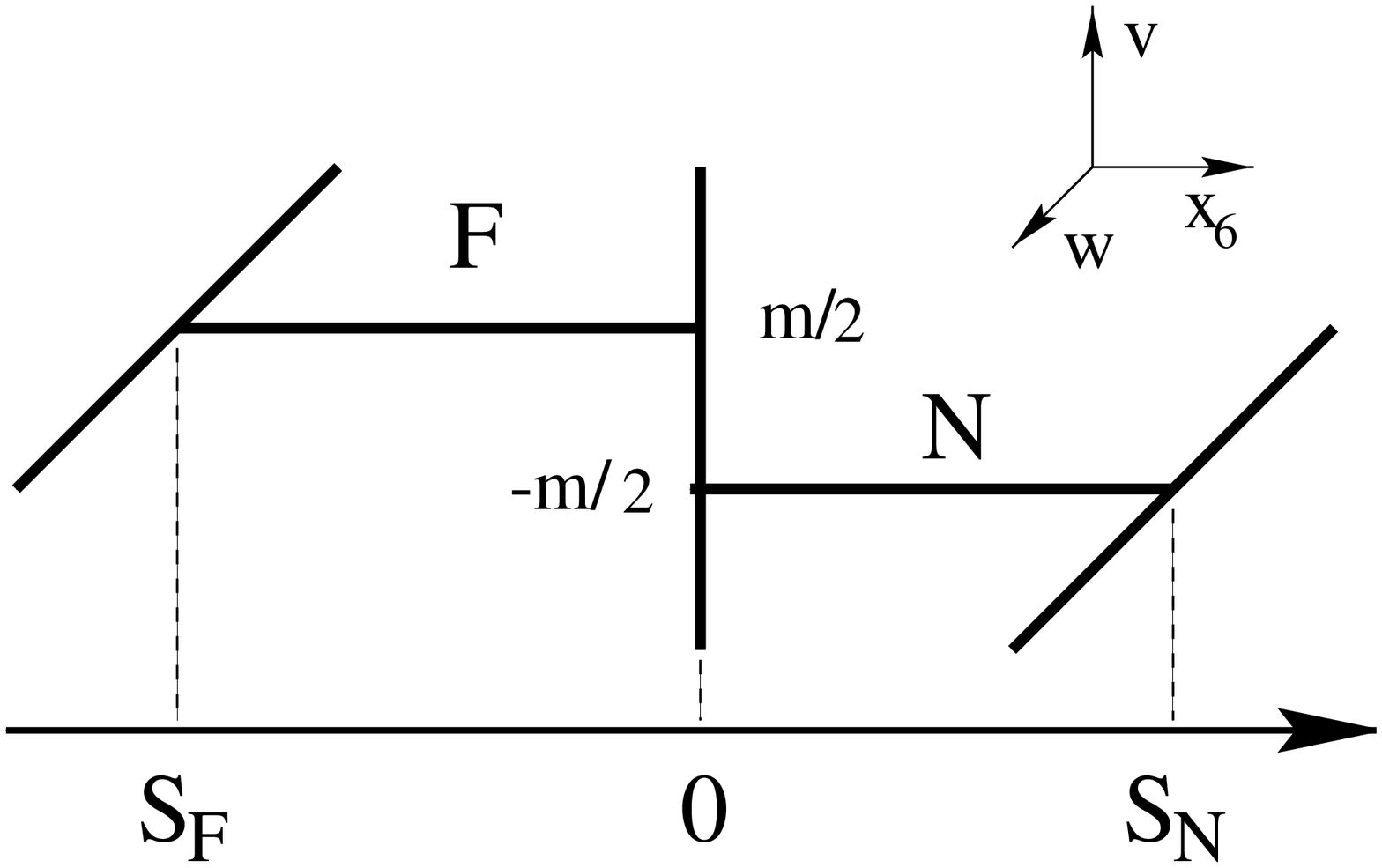}}
\vskip.2in
\noindent
Fig 6. {\it The classical string theory brane configuration corresponding
to SQCD with a gauged flavor group and a mass $m$ for the quarks.
The NS5 brane extending into the $v$ direction is located at $x_6=0$
whereas the NS5 branes extending into the
$w$ direction are at $x_6=S_F$ and $x_6=S_N$.}
\vskip .2in
\end{figure}

In M-theory the brane configuration  corresponding to the string
theory picture of Fig. 6. is a single 5-brane  given by the following curve,
\beq
t &=& m^{F-N} {(v+m/2)^N \over (v-m/2)^F} \\
w &=& m\left[{(m^{F-N}\LaN^{3N-F})^{1/N} \over v+m/2}
+{(m^{N-F}\LaF^{3F-N})^{1/F} \over v-m/2} \right]\ .
\eqn{FNcurve}
\eeq

Again we ignore a $g_s$-dependent factor
in the $w$ equation.

We can check that the curve has the correct
behavior at its infinities
\beq
\eqn{NFinf}
i. \quad v \to {m \over 2} \quad & t \to \infty, w \to \infty \quad
 &t \sim {w^F\over m^{N-F}\LaF^{3F-N}}, \nonumber\\
ii. \quad v \to -{m \over 2} \quad & t \to 0, w \to \infty \quad
 &t \sim (-1)^F {m^{F-N}\LaN^{3N-F} \over w^N}, \nonumber\\
iii. \quad v \to \infty \quad & t \to 0, w \to 0 \quad
 &t \sim \left({m\over v}\right)^{F-N}\ .
\eeq
One can check that the curve has
the spurious $R_v$ and $R_w$ symmetries of the field theory
\eq{NFtable}, that the symmetries are broken to the correct
discrete subgroups at the infinities of the curve as
expected from instantons in the field theory,
and that finally all symmetries are broken by the whole
curve when the spurious transformations are turned off.

As in the case of semi-infinite branes the limit of weakly coupled
string theory is reached by taking $R\to 0$ with
$\LaF,\LaN \to 0$. If we take
\beq
\eqn{Mscales}
\LaN^{3N-F}=e^{-S_N/R}\ ,
\qquad \LaF^{3F-N}=e^{S_F/R}
\eeq
where $S_N>0$ and $S_F<0$ as well as $m$ are held fixed, we reproduce
Fig. 6 with the NS 5 branes positioned at $x_6=S_F,0,S_N$
in the limit $R \to 0$.

\subsection{Duality}

Duality for the string theory setup of Fig. 6. might at first seem
a little puzzling. From the field theory we know of the existence
of a dual with gauge group $SU(F-N)$ which we might expect to be
able to reproduce in string theory by moving the rightmost NS 5-brane
 to the left of the NS 5-brane at $x_6=0$.
From arguments based on D4-brane charge conservation on the
NS 5-branes one can deduce that the final arrangement has to
have $F-N$ D4-branes suspended between the two branes on the
right and $F$ D4-branes on the left. The puzzle is that the $F$
D4-branes on the left are suspended between two parallel NS 5-branes
 which live at different values of $v$. But this configuration with D4-branes
at an angle breaks the remaining supersymmetry.

The resolution of the puzzle is quite simple. We know from the field
theory that the dual gauge theory breaks supersymmetry
classically, and supersymmetry is 
only restored through a nonperturbative quantum
effect. The string theory brane configuration
exactly reproduces the classical result. To see the corresponding
quantum effect we should turn to the M theory curve
which -- because it is defined by the complex curve of eqs. (6.5, 6.6)
 -- preserves supersymmetry.

Recall from the discussion of SQCD with semi-infinite branes
that we can find the string theory setup of the dual theory
by a smooth deformation of the M-theory curve which corresponds to
taking  $\LaN \gg 1$. 
From \eq{Mscales} we see that this implies that $S_N$ is negative.
Then we take the limit of $R \to 0$ while holding $S_N$, $S_F$ and
the meson vev fixed.
This also determines the required scaling of the masses
\beq
m^{F-N}=e^{S_N/R}\ ,
\eeq
and we see that the masses are driven to zero in the classical limit.
Thus the M-theory limit shows that the classical brane configuration
which preserves supersymmetry necessarily has zero masses. For
any non-zero mass the D4-branes at an angle break supersymmetry.
However, in the quantum theory a nonperturbative effect allows the D4-branes
 to bend while preserving supersymmetry. This conclusion
can be supported by calculating the non-perturbative superpotential
of the strong field theory gauge dynamics directly from the M-theory
curve using an expression for the superpotential recently introduced by 
Witten \cite{witten2},  given by an integral of the
holomorphic 3-form over the curve.
We have checked that a straightforward but tedious generalization of
Witten's calculation
to the case at hand 
reproduces the field theory result \eq{wdynFN}.

\begin{figure}[t]
\centerline{\epsfxsize=4 in \epsfbox{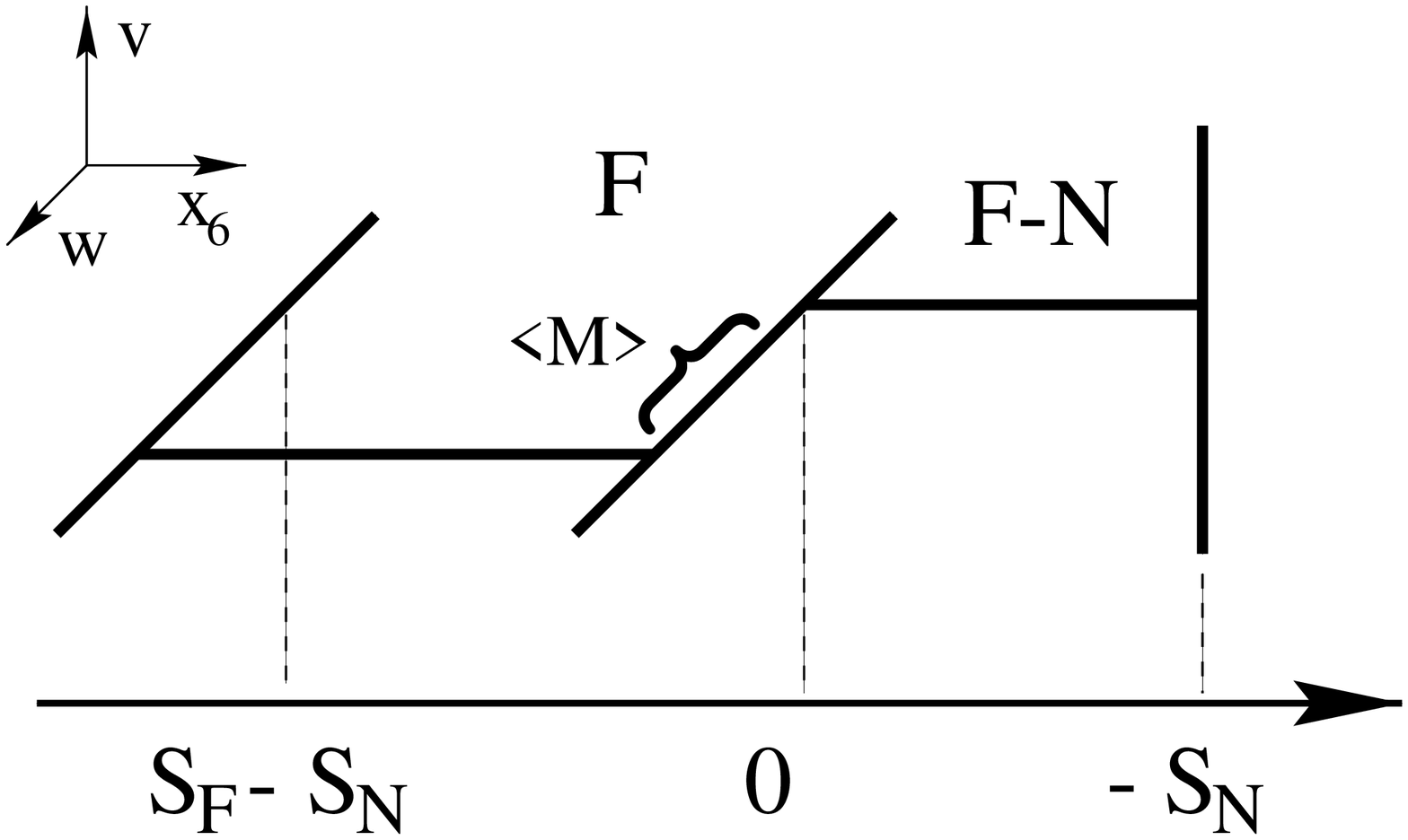}}
\vskip.2in
\noindent
Fig 7. {\it The classical string theory brane configuration corresponding
to the dual of SQCD with a gauged flavor group and an expectation value
for the meson field.}
\vskip .2in
\end{figure}

The low energy particle content of the dual theory can now
be read off as short strings connecting the various D4-branes.
The $SU(F-N)$ gauge bosons arise from strings between the
$F-N$ color branes on the right of Fig. 7, and the $F$ flavors of
dual quarks are strings connecting color and flavor branes.
Since the two NS 5-branes on the left of the diagram are now parallel,
the flavor branes suspended between them can slide freely in the
$w$ direction. The coordinates of the D4-branes in this direction
are interpreted as  the meson chiral
superfield. Another way of seeing the existence
of the meson is to note that the flavor sector is $N=2$ supersymmetric.
Therefore the spectrum of strings connecting the flavor branes not
only gives an $N=1$ gauge multiplet of the $SU(F)$ flavor group but
there must also be 
an adjoint chiral superfield,  $M_{\rm adj}$. This adjoint is coupled to
the quark fields in the $N=2$ symmetric superpotential term
$W=\twi Q_+ M_{\rm adj} \twi Q_-$ as required by the field theory
duality.

Two issues in the above story need further clarification.
Firstly, we have  treated the flavor group as a very
weakly coupled spectator
gauge group and have argued that we can ignore the dynamics of
this group. But the gauge coupling of the $SU(F)$ is related
to the Yukawa coupling of the meson in the dual by the approximate $N=2$
supersymmetry of the flavor sector. Thus, if we take the gauge coupling
small we are forced to also take the Yukawa coupling small.
However this is not a problem since the $N=1$ duality is only expected
to hold at the infrared  fixed point.
For example, if we take $3N/2 < F <3N$
so that the infrared regime of SQCD is conformal, then the dual
$SU(F-N)$ gauge coupling will also approach a fixed point with
associated anomalous dimensions for the quarks (which break the 
approximate $N=2$
supersymmetry badly) and  the Yukawa coupling will increase to the
SQCD fixed point value. These anomalous dimensions 
 will also contribute to the running of the
asymptotically free $SU(F)$ coupling but in the limit
$\LaF \ll m \ll \LaN$ that we are considering the theory can be made
to approach the fixed point arbitrarily closely before the $SU(F)$
dynamics becomes important.

Secondly, we seem to be missing the trace of the meson field
in the spectrum of our dual. This field is the superpartner
of the diagonal $U(1)\subset U(F)$, which is ``frozen'' by the
quantum bending of the NS 5-branes. We believe that the
missing singlet might be recovered by considering the
above descibed brane setup with an additional D4-brane
suspended between the two NS 5-branes which extend to infinity
in $w$. This additional brane does not change the perturbative spectrum
and breaks supersymmetry if the non-perturbative dynamics is ignored.
In the full quantum theory its presence has the effect of ``thawing''
the diagonal $U(1)$ and its scalar partner by eliminating the
infrared divergence which ``froze'' it \cite{witten1,egkrs}.

\section{Conclusions}

We have derived Seiberg's duality for $N=1$ supersymmetric QCD from
simple M-theory considerations. We expect that this M-theory approach to
$N=1$ duality can be generalized to more complicated theories. It may
thereby provide a better understanding of the unifying principle behind
duality, as well as the connection between electric and magnetic dual
pairs. Our work elucidated such phenomena as the non-perturbative
bending of D4-branes and the creation of D4-branes as NS 5-branes are
passed through each other. More work needs to be done to satisfactorily
resolve the puzzle of the missing flavor singlet meson in the magnetic
dual theory arising from our M-theory considerations. It is possible
that this may become clearer by studying configurations including
$6$-branes.
   
\section*{Acknowledgements}

This research was supported by the U.S. Department of Energy under grant
\# DE-FG02-94ER40818.
We are grateful to Andrew Cohen, Csaba Csaki, Zachary Guralnik, Witold
Skiba, and especially Nick Evans for 
discussions on topics related to this paper.


\end{document}